\documentclass[journal,twoside,web]{ieeecolor}
\usepackage{tmi}
\usepackage{cite}
\usepackage{amsmath,amssymb,amsfonts}
\usepackage{algorithmic}
\usepackage{graphicx}
\usepackage{textcomp}
\usepackage[detect-all=true,separate-uncertainty=true, multi-part-units=single,range-phrase={-},product-units = single]{siunitx}
\usepackage{menukeys}
\usepackage{amsfonts}
\usepackage{import}
\usepackage{booktabs}
\usepackage{xcolor}
\usepackage[caption=false,font=small]{subfig}
\usepackage{wasysym}
\usepackage{booktabs}
\usepackage{dcolumn}
\usepackage{chemformula}
\usetikzlibrary{calc, arrows, positioning, shapes.geometric, intersections, fit, math}

\usepackage[switch]{lineno}

\definecolor{myblue}{RGB}{0,0,128}
\def\BibTeX{{\rm B\kern-.05em{\sc i\kern-.025em b}\kern-.08em
    T\kern-.1667em\lower.7ex\hbox{E}\kern-.125emX}}
    
\begin{document}
\bstctlcite{IEEEexample:BSTcontrol}
\title{Deep Learning for High Speed Optical Coherence Elastography with a\\ Fiber Scanning Endoscope}
\author{Maximilian~Neidhardt, Sarah~Latus, Tim~Eixmann, Gereon H\"uttmann and Alexander~Schlaefer
\thanks{This work was partially funded by the TUHH $i^{3}$ initiative, the NaviNad project (DFG, grant SCHL 1844-2-2), the Interdisciplinary Competence Center for Interface Research (ICCIR) by UKE and TUHH, and the European Union under Horizon Europe programme (No. 101059903).}\thanks{M. Neidhardt, S. Latus, and A. Schlaefer are with the Institute of Medical Technology and Intelligent Systems, Hamburg University of Technology, Hamburg, Germany.}\thanks{T. Eixmann and G. H\"uttmann are with the Institute of Biomedical Optics, University of L\"ubeck, L\"ubeck, Germany.}\thanks{A. Schlaefer is with the SustAInLivWork Center of Excellence}\thanks{M. Neidhardt and A. Schlaefer are with the ICCIR.}}

\maketitle

\begin{abstract}
Tissue stiffness is related to soft tissue pathologies and can be assessed through palpation or via clinical imaging systems, e.g., ultrasound or magnetic resonance imaging. Typically, the image based approaches are not suitable during interventions, particularly for minimally invasive surgery. To this end, we present a miniaturized fiber scanning endoscope for fast and localized elastography. Moreover, we propose a deep learning based signal processing pipeline to account for the intricate data and the need for real-time estimates. Our elasticity estimation approach is based on imaging complex and diffuse wave fields that encompass multiple wave frequencies and propagate in various directions. We optimize the probe design to enable different scan patterns. To maximize temporal sampling while maintaining three-dimensional information we define a scan pattern in a conical shape with a temporal frequency of \SI{5.05}{\kilo\hertz}. To efficiently process the image sequences of complex wave fields we consider a spatio-temporal deep learning network. We train the network in an end-to-end fashion on measurements from phantoms representing multiple elasticities. The network is used to obtain localized and robust elasticity estimates, allowing to create elasticity maps in real-time. For 2D scanning, our approach results in a mean absolute error of \SI{6.31(576)}{\kilo\pascal} compared to \SI{11.33(12.78)}{\kilo\pascal} for conventional phase tracking. For scanning without estimating the wave direction, the novel 3D method reduces the error to  \SI{4.48(3.63)}{\kilo\pascal} compared to \SI{19.75(21.82)}{\kilo\pascal} for the conventional 2D method. Finally, we demonstrate feasibility of elasticity estimates in ex-vivo porcine tissue. 
\end{abstract}


\section{Introduction}
\IEEEPARstart{Q}{uantitative} estimates of soft tissue elasticity can be achieved using shear wave elastography imaging. First, a disturbance is excited inside the tissue and second, the propagating waves are tracked with a high-frequency imaging modality. The estimated phase velocity of a wave field can thereby be directly related to the elasticity of soft tissue. The principle has been intensively studied with ultrasound imaging or magnetic resonance imaging (MRI)~\cite{scott2020artificial} and demonstrated in clinical studies, e.g., for staging liver fibrosis~\cite{Lee.2017}. Still, MRI is \textcolor{black}{challenging during minimally invasive interventions} and ultrasound signals can be obstructed by bone or dense soft tissues, and imaging deep tissues is limited due to the damping of the signal, especially in morbidly obese patients~\cite{Lee.2017}. Therefore, current imaging methods are unsuitable for estimating quantitative elasticity values during interventions, especially in the context of minimally invasive surgery.

\begin{figure}[tb!]
        \includegraphics[width=\linewidth]{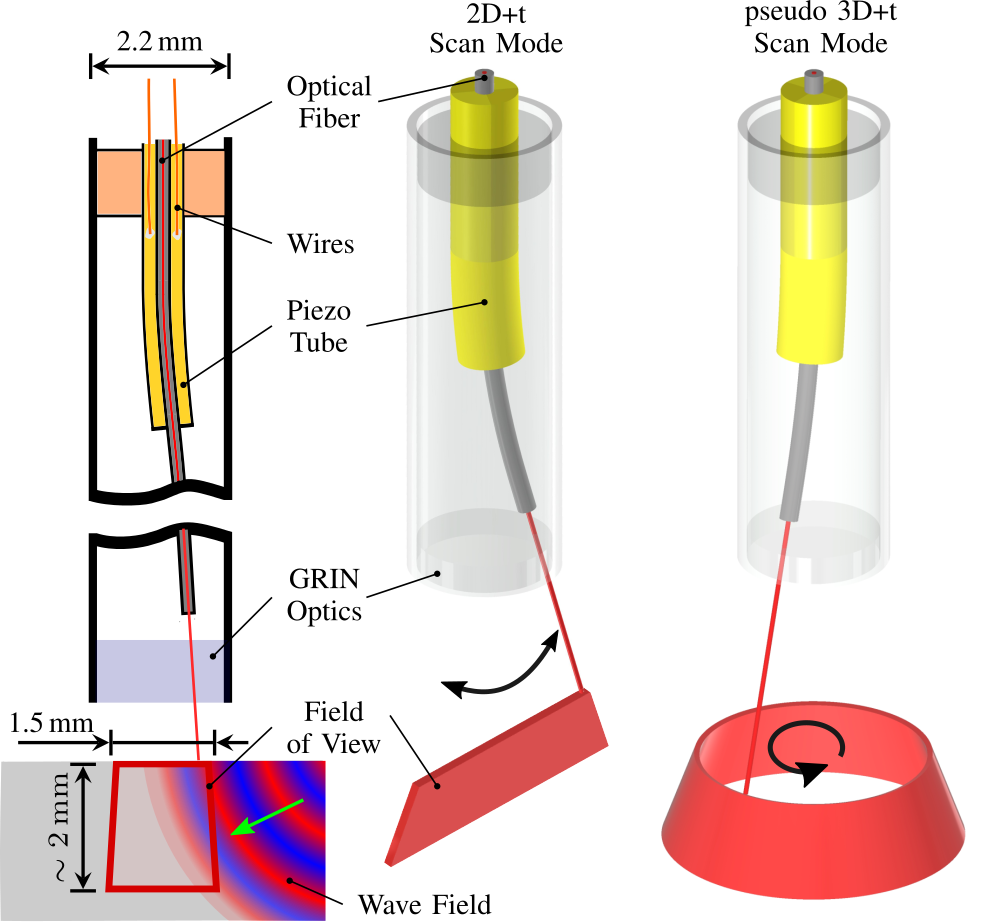}
        \caption{\textbf{Optical Fiber Scanning Endoscope} for data acquisition. We acquire image data with a high temporal frequency of a propagating wave field. The imaging fiber is positioned inside a piezo tube which is deflected by sinusoidal amplified signals for data acquisition in 2D+t or pseudo 3D+t scan mode.}
        \label{fig:FSE}
\end{figure}

Optical coherence tomography (OCT) offers imaging of tissue with a high axial and temporal resolution. OCT imaging has been applied during minimally invasive surgery, e.g., imaging brain tissue~\cite{finke2012automatic} or catheter-based interventions~\cite{latus2019bimodal, gessert2018automatic}. Usually, galvanometer scanners are used to deflect the one-dimensional OCT beam, allowing the acquisition of two- or three-dimensional images. Recently, high-speed swept-source OCT systems (SS-OCT) have been proposed for tracking rapidly changing wave fields in two-~\cite{Manapuram.2012, Neidhardt.2020} or three-dimensional images over time~\cite{Neidhardt.2021}. These systems have been successfully applied in estimating the elasticity of superficial structures, e.g., in ophthalmology~\cite{Tao.2014}. Still, imaging depth in tissue is limited to a few millimeters with OCT, and hence a small OCE probe that can be used during minimally invasive surgery inside the patient would be desirable.

In OCT imaging, light travels in optical fibers that have a diameter of approximately \SI{100}{\micro\meter} allowing the integration into miniaturized imaging probes. It has been shown that needle-based designs can estimate elasticity based on OCT imaging of the compressed tissue in front of the needle~\cite{Ejofodomi.2013, Qiu.2016, Wang.2022b}. Thereby, the speckle movement is tracked and a material model is derived which maps speckle translation to qualitative elasticity estimates. Typically, waves are excited inside the tissue by an ultrasound acoustic radiation force impulse\cite{Latus2017, neidhardt2022ultrasound} or externally by a vibration source, e.g., an air pulse~\cite{wang2013focused} or a vibrator~\cite{liu2021two}. Quantitative shear wave elastography approaches with miniaturized probes have been presented with one-dimensional depth scans and external excitation~\cite{Latus2017}.

Several groups have also shown that wave excitation can be integrated into a probe using miniaturized ultrasonic elements~\cite{Karpiouk.2018, Qu.2017} or piezoelectric actuators~\cite{Parmar.2021}. For these transient elastography approaches, the spatial distance and time between excitation and detection of the wave are calculated to estimate the phase velocity. Hence, calibration of spatial distances between excitation and imaging field of view (FOV) as well as an accurate triggering is crucial. Dual-fiber probe designs can improve the sampling of the shear wave propagation by minimizing the influences of the distance and orientation to the excitation source~\cite{Latus.2023}. However, a two- or three-dimensional imaging of rapidly changing wave fields is not possible with all of these miniaturized probes. Recently, fiber scanning endoscopes (FSE) have made it possible to perform versatile scan patterns to capture two- or three-dimensional OCT images. The optical fiber is integrated into a piezoelectric tube~\cite{Huo.2010, Liang.2017, SchulzHildebrandt.2018} or magnetic driven cantilevers~\cite{Huang.2011}. 
Applications of these probes to assess the tissue morphology, angiography, and strain-based elastography~\cite{nakamura2012endoscopic} have been proposed. However, a FSE for quantitative shear wave elastography has not yet been presented. Tracking of propagating waves requires a multi-dimensional sampling with a high temporal frequency that has not been achieved with current FSE probe designs. 
In this work, we present the application of a bespoke FSE with an outer diameter of \SI{2.2}{\milli\meter} for minimally invasive elasticity imaging. We design our probe to achieve real-time imaging of rapidly changing wave fields. We optimize our probe design to perform robust line (2D+t) or conical shape scanning patterns (pseudo 3D+t) with imaging frequencies of up to \SI{10.1}{\kilo\Hz}, as depicted in Figure~\ref{fig:FSE}.
Extracting shear wave properties from complex scan patterns in real time is challenging. Therefore, we propose deep learning to estimate the elasticity of tissue directly from the complex OCT image data. During imaging, we continuously excite multi-frequency waves with a piezoelectric transducer positioned at the surface of the tissue. The setup does not require triggered events, as we aim to capture a complete scan pattern sequence of the wave fields. We compare the accuracy of deep learning based elasticity estimates to a conventional approach that estimates the phase velocities from two-dimensional images in the frequency domain~\cite{Maksuti.2016, Beuve.2021b}. We propose deep learning to infer quantitative elasticity values directly from sparsely sampled pseudo 3D+t scan patterns. Deep learning approaches have been presented for highly sampled image data acquired with OCT scan apertures in 2D+t~\cite{Neidhardt.2020} and 3D+t~\cite{Neidhardt.2021}, and with ultrasound imaging~\cite{ vasconcelos2021viscoelastic, neidhardt2022ultrasound, yin2023swenet}. Typically, inside the patient the position of the FSE relative to the excitation source is unknown. Hence, we systematically analyze the elasticity estimates concerning the imaging position relative to the excitation location with an experimental setup including a robot on soft tissue phantoms and ex-vivo soft tissue samples. Our data processing approach demonstrates that a robust, local, contactless, and non-directional estimate of tissue elasticity is feasible with our FSE.

\section{Material and Methods}
\begin{figure}[b!]
    \centering
    \includegraphics[width=\linewidth]{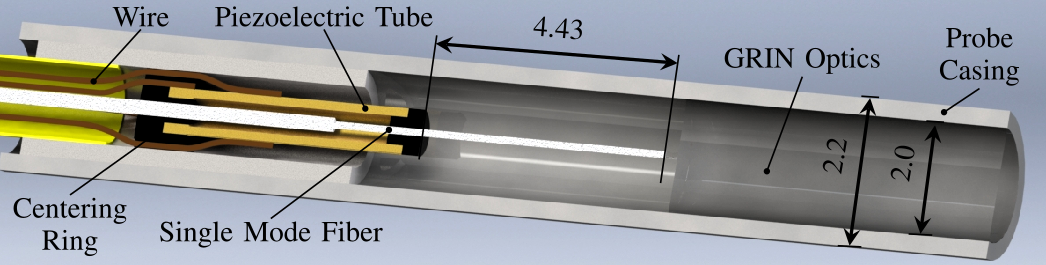};

    \caption{\textbf{CAD Design:} Cross-section view of the individual components of the fiber scanning endoscope. All measurements are given in millimeters.}
    \label{fig:FSE_CAD}
\end{figure}

\begin{figure*}[tb!]
    \centering
         \includegraphics[width=\linewidth]{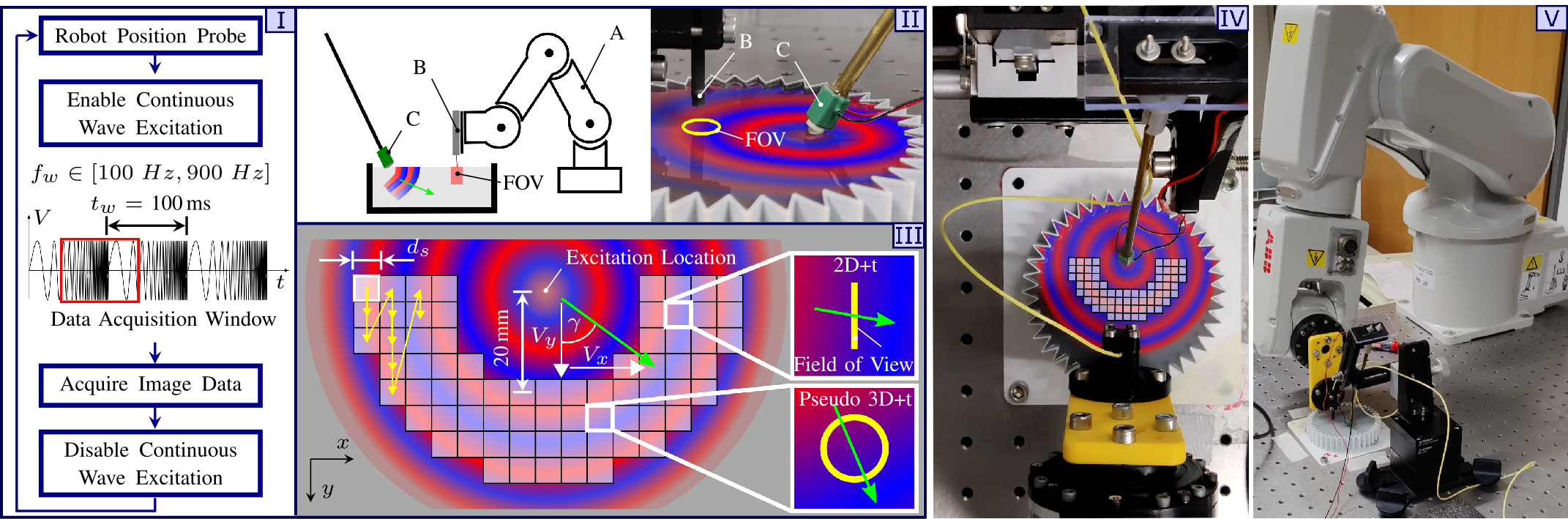}
    
    \caption{\textbf{Experimental Setup for Data Acquisition:} \textbf{[I]} Flow chart with data acquisition steps. \textbf{[II]} Side-view of the experimental setup with a robot (A), FSE (B), and the piezo element for continuous exciting of a wave field (C).\textbf{ [III]} Top view of gelatin phantom; the spherical wave field is indicated in red and blue. The green arrow indicates the direction of the propagating waves and parts of the trajectory of the robot are depicted in yellow. \textbf{[IV]} Top view and \textbf{[V]} overall experimental setup in the lab excluding the OCT imaging system.}
    \label{fig:expsetup}
\end{figure*}

\subsection{Fiber Scanning Endoscope}
For image data acquisition of a diffuse wave field, we \textcolor{black}{designed} a bespoke FSE with an outer diameter of \SI{2.2}{\milli\meter} which can be inserted into a conventional clinical laparoscopic trocar. The optical design is based on a single mode fiber mounted inside a piezoelectric tube as depicted in Figure~\ref{fig:FSE_CAD}. \textcolor{black}{The eigenmodes of the free end of the fiber were simulated (COMSOL Multiphysics, Stockholm, Sweden) and following we fabricated} our FSE with a free end of \SI{4.43}{\milli\meter}. The light ray is focused by a Gradient-index (GRIN) optics (GT-IFRL-200-005-50C, Grintech GmbH, Germany) located at the outer tubing close to the probe end facet. We report an ideal working distance of \SI{3.5}{\milli\meter} and a lateral resolution of \SI{16}{\micro\meter} FWHM. The piezoelectric tube is divided into four sections which allow a deflection along both lateral axes. The probe is driven by two sinusoidal signals with a frequency of \SI{5.05}{\kilo\hertz}, a phase shift of \SI{90}{\degree} and a voltage of \SI{200}{Vpp}. The signals \textcolor{black}{were} generated in LabVIEW (Spring2017, National Instruments, Austin, Texas, U.S.) with a connected DAQboard (USB-6343, National Instruments, Austin, Texas, U.S.) and \textcolor{black}{were} fed into a multi-channel high voltage amplifier (E-413.6, Physik Instrumente, Lederhose, Germany). The FSE can be operated in either line or conical shape scan mode to acquire 2D+t or pseudo 3D+t data representations, respectively. In 2D+t scan mode, cross-sectional images \textcolor{black}{were} acquired with a temporal sampling rate of \SI{10.1}{\kilo\hertz} and dimensions of approximately \qtyproduct{1.5 x 2}{\milli\meter} along the lateral and depth axis, respectively. In pseudo 3D+t scan mode, the fiber is deflected along both lateral axes resulting in a conical shape scanning trajectory as depicted in Figure~\ref{fig:expsetup} (II). The diameter of the circle is approximately \SI{1.5}{\milli\meter} and a temporal sampling rate of \SI{5.05}{\kilo\hertz} is achieved. Please note, that the trajectories contain imperfections due to inaccuracies during the manufacturing of the probe in our lab. These can be partially calibrated and applied during image reconstruction~\cite{SchulzHildebrandt.2018}. However, in this work, all spatial calibrations of the FSE are included in the deep learning model.

\subsection{Experimental Setup and Data Acquisitions}
The experimental setup is depicted in Figure~\ref{fig:expsetup} (II-V). We use a high-speed swept-source OCT system (SS-OCT, OMES, Optores, Germany) with a temporal scan rate of \SI{1.5}{\mega\Hz}, a central wavelength of \SI{1315}{\nano\meter} and an axial resolution of \SI{15}{\micro\meter} in air. We \textcolor{black}{designed} the FSE to operate at a cycling rate of \SI{5.0516}{\kilo\Hz}. Considering the temporal scan rate of the OCT system this leads to 314 scan lines per scan cycle. A piezoelectric element (Piezomechnik GmbH, Germany) with dimensions \qtyproduct{7 x 7 x 5}{\milli\meter} \textcolor{black}{was} fitted with an epoxy dome to facilitate contact to soft tissue and \textcolor{black}{was} positioned on the surface of the phantom to excite a wave field continuously. For multi-frequency actuation of the piezoelectric element, we \textcolor{black}{applied} a frequency $f_{w} \in [100~Hz, 900~Hz]$ with a swept time $t_w$ of \SI{100}{\milli\second}, and a peak-to-peak voltage of \SI{120}{Vpp}. \textcolor{black}{The FOV of the FSE was estimated} in the 2D+t scan mode with a calibration phantom (R3L3S3P, Thorlabs, Germany). For positioning the endoscope relative to the phantom we use a robot arm (IRB 120, ABB) with a high positioning repeatability of \SI{0.01}{\milli\meter}. We \textcolor{black}{mapped} elasticities with a scan field consisting of squares with dimension \qtyproduct[parse-numbers = false]{d_s x d_s}{}, as depicted in Figure~\ref{fig:expsetup} (III). The robot \textcolor{black}{positioned} the FSE at the center of each square.

 \begin{figure*}[t]
        \subfloat[Pseudo 3D+t Scan Mode]{
        \includegraphics[height=60mm]{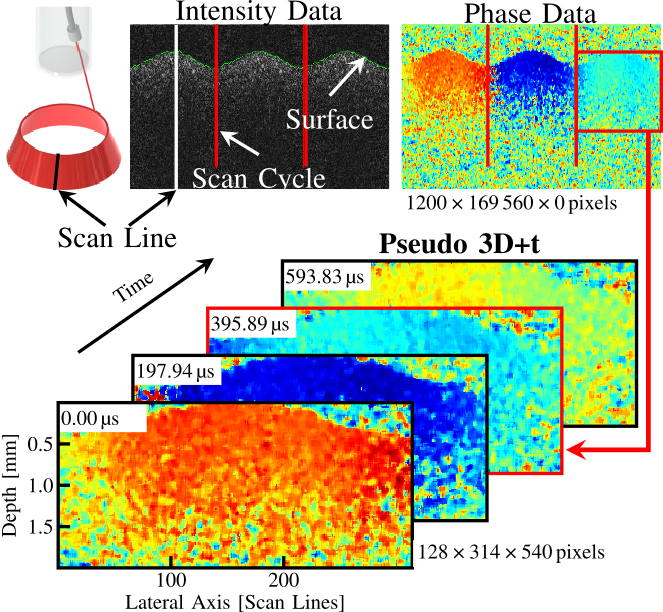}
           \label{fig:dataRepresentations_3D_t}
    }
    \hfill
    \subfloat[2D+t Scan Mode]{
        \includegraphics[height=60mm]{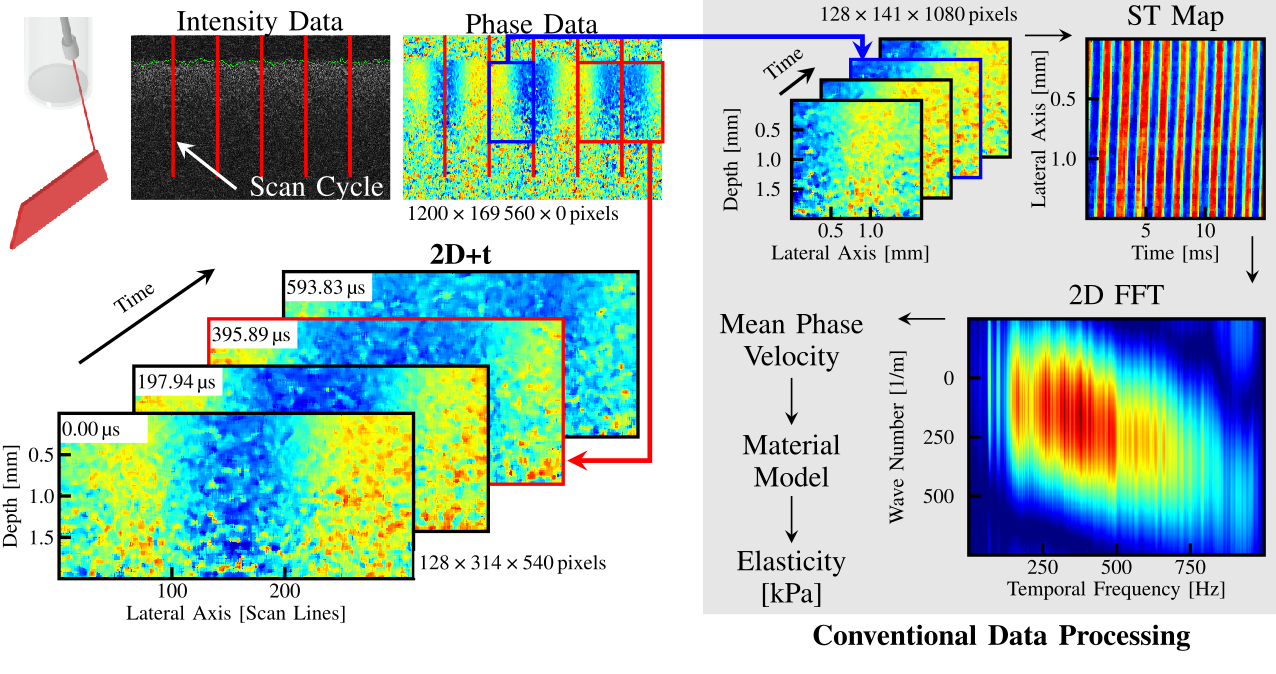}
        \label{fig:dataRepresentations_2D_t}
    }    
    \caption{\textbf{Data Representations:} We image propagating waves with our optical fiber endoscope and acquire (a) 2D+t data or (b) pseudo 3D+t data sets. For conventional image data processing, we estimate a space-time map (ST Map) from 2D+t data sets. We perform a 2D FFT and calculate the mean phase velocity. We map the estimated velocity to elasticity values with a material model.}
    \label{fig:dataRepresentations}
\end{figure*}

Figure~\ref{fig:expsetup} (I) depicts the individual steps performed during data acquisition. First, we \textcolor{black}{placed} the piezoelectric element on the surface of the phantom using a motion stage and the robot positions the FSE. Next, we continuously \textcolor{black}{excited} a multi-frequency wave field inside the phantom with the piezoelectric element and simultaneously \textcolor{black}{acquired} image data with the FSE. Please note, that we \textcolor{black}{acquired} image data sets without any triggered events. Each data set contains \num{169560} scan lines with a resolution of 600 pixels along the depth axis and a recording time of \SI{106.89}{\milli\second}. This guarantees us, that all excited waveforms \textcolor{black}{were} included in a data set. The individual positions for data acquisition and the trajectory of the robot are depicted in Figure~\ref{fig:expsetup} (III). In total, we \textcolor{black}{acquired} image data sets at 70 positions. We \textcolor{black}{repeated} this procedure for both scanning modes.

\subsection{Data Representation and Preprocessing}
The individual steps are depicted in Figure~\ref{fig:dataRepresentations}. For 2D+t data sets, we initially \textcolor{black}{estimated} the turning points of the FSE in the intensity OCT data by iteratively removing scan lines from the beginning and applying crops of 314~pixels along the temporal axis. This \textcolor{black}{allowed} us to estimate the minimum difference in intensity of two subsequent crops, thereby determining the turning points during an imaging cycle of the FSE. Pseudo 3D+t data sets \textcolor{black}{were} simply cropped at multiples of 314~scan lines, as no turning point is given. The following preprocessing steps \textcolor{black}{were} applied to 2D+t and pseudo 3D+t data sets: First, we \textcolor{black}{extracted} scan lines in the phase part of the complex-valued OCT signal that \textcolor{black}{were} acquired at the same spatial position during a scan. We \textcolor{black}{unwraped} these scan lines along the temporal axis and \textcolor{black}{estimated} the phase difference between subsequent scan lines. Second, we \textcolor{black}{detected} the surface of the sample in the intensity data and \textcolor{black}{cropped} sequences of \SI{128}{pixels} along the depth axis and 314 scan lines. We set the surface topmost pixel of the data set as the starting point along the depth axis. We report for 2D+t and 3D+t data sets a size of \qtyproduct{128 x 314 x 540}{pixels} along the depth, lateral and time axis, respectively.
For conventional phase velocity estimation we \textcolor{black}{created} space-time (ST) maps from 2D+t data sets. We \textcolor{black}{cropped} sequences of 157 scan lines from the phase data, as indicated by the blue box in Figure~\ref{fig:dataRepresentations_2D_t}. We \textcolor{black}{flipped} every second frame along the temporal axis, remove scan lines in the proximity of the turning point, and report a data set size of \qtyproduct{128 x 141 x 1080}{pixels}. ST Maps \textcolor{black}{were} generated by computing the mean along the depth axis resulting in an image resolution of \qtyproduct{141 x 1080}{pixels} along the lateral and time axis, respectively.

\begin{figure*}[ht]
    
    \includegraphics[width=\linewidth, trim={0 0 0cm 0},clip]{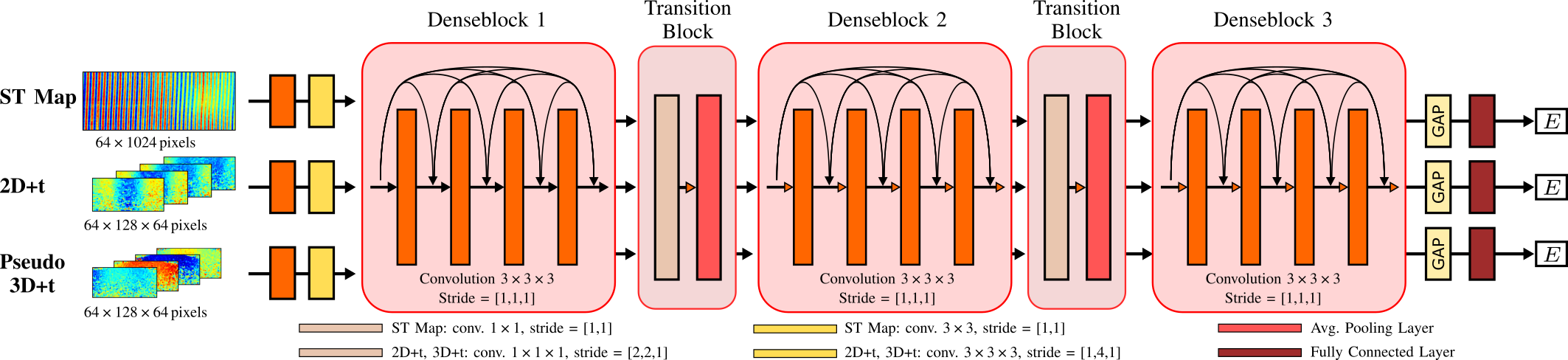};
    
    \caption{\textbf{Spatio-Temporal DenseNet:} We predict the elasticity from ST Maps, 2D+t or pseudo 3D+t phase data sets. For 2D+t and 3D+t data representations we crop sequences along the temporal axis. The input data set sizes are indicated on the left along the depth, lateral, and temporal axis, respectively. The architecture consists of an initial part with an average pooling layer and a convolution layer, followed by three DenseNet blocks, which are connected with transition layers. The last block is connected to a global average pooling (GAP) layer, and the output is fed into the regression output layer. Please note, that we train individual networks for each data structure.}
    \label{fig:DenseNet}
\end{figure*}

\subsection{Conventional Phase Velocity Estimation}
The data processing for our conventional phase velocity estimation is depicted in Figure~\ref{fig:dataRepresentations_2D_t}, right. The phase velocity \textcolor{black}{was} estimated from ST Maps similar to~\cite{Maksuti.2016,Beuve.2021b}. We \textcolor{black}{transformed} the ST Map into the $k$-space by using the 2D discrete FFT. To reduce imaging noise, a band-pass filter with a center frequency of \SI{400}{\hertz} and a band size of \SI{200}{\hertz} \textcolor{black}{was} applied. To further reduce background noise, we \textcolor{black}{removed} pixels with less than $10\%$ of the overall maximum amplitude in the k-space, similar to~\cite{Maksuti.2016}. We \textcolor{black}{estimated} the maximum amplitude for each temporal frequency $f$ and \textcolor{black}{computed} the phase velocity $v_{FFT} = {f}/{k}$ with the wavenumber $k$. Estimates in the range $v_{FFT}>\SI[]{10}{\meter\per \second}$ are considered failed estimates, which is above the typical elasticity range of soft tissue~\cite{fink2010multiwave} and are depicted in elasticity maps in gray.
While the robot positions the FSE, the direction of the wave propagating through the FOV changes. We define the angle $\gamma$ between the lateral orientation of the image and the propagating direction of the wave as $\tan(\gamma) = v_{x}/v_{y}$ with the velocity components $V_{x}$ and $V_{y}$ as indicated in Figure~\ref{fig:expsetup} (III). The lateral imaging axis during 2D+t data acquisition \textcolor{black}{was} throughout data acquisition aligned with the robot's coordinate system. For conventional phase velocity estimates, we assume that the position of the FSE and the piezoelectric element \textcolor{black}{was} known. Thereby, we can correct estimated velocities $v_{FFT}$ depending on the position of the probe by $v_{FFT_{AC}} = \sin(\gamma)*v_{FFT}$ and refer to this data set as FFT + AC. Exclusively for the conventional approach, we \textcolor{black}{mapped} the estimated velocities $v$ to the Young's Modulus $E$ by a linear regression approach defined as $E=k\cdot v +q$. We \textcolor{black}{minimized} the absolute error between estimates and define $k= 24.2$ and $q=-16.4$. This allows a fair comparison to our deep learning approach. Please note, that the position of the FSE relative to the piezoelectric element \textcolor{black}{was} only utilized for conventional data processing while our deep learning data processing purely \textcolor{black}{relied} on the acquired image data.\\

\vspace{-3mm}

\subsection{Deep Learning Model}
We estimate elasticity with spatio-temporal convolutional neural networks, which jointly learn the spatial and temporal image features. As a baseline, we \textcolor{black}{used} densely connected neural networks~\cite{huang2017densely}. Spatio-temporal networks have shown promising results for predicting phantom stiffness due to their parameter and computational efficiency~\cite{Neidhardt.2020, Neidhardt.2021, neidhardt2022ultrasound}. We \textcolor{black}{designed} a custom DenseNet architecture as depicted in Figure~\ref{fig:DenseNet}. The network consists of three main DenseNet blocks with a kernel size of three in all dimensions and a stride of one. Transition blocks connect the DenseNet blocks and contain an average pooling layer for downsampling of the input size. For ST Map and 2D+t data representations, we \textcolor{black}{applied} a stride of one in each dimension. For \textcolor{black}{the pseudo 3D+t} data representation, we \textcolor{black}{applied} a stride of two along the depth and lateral axis and one along the temporal axis, to increase efficiency during training. Throughout the network we do not downsample features along the temporal axis to preserve a high temporal sampling resolution. This is crucial for tracking rapidly changing wave fields. In contrast, we \textcolor{black}{applied} a coarse filter along the spatial dimensions \textcolor{black}{to incorporate the spatial dependencies of the FSE in the network}. Zero padding \textcolor{black}{was} applied to preserve the input size throughout the convolutional layers. For all convolutional layers we \textcolor{black}{used} batch normalization~\cite{batchnorm2015} and the rectified linear activation function for our network layers. The network contains a total of \num{109690} trainable parameters \textcolor{black}{and is real-valued}. The last DenseNet block is connected to a global average pooling (GAP) layer followed by a fully connected layer which outputs the Young's Modulus $E$. All implementations \textcolor{black}{were} performed with \textcolor{black}{\textit{PyTorch-Lightning}} \cite{Falcon_PyTorch_Lightning_2019} and the training \textcolor{black}{was} executed \textcolor{black}{on a GPU (GeForce RTX 4090, NVIDIA, U.S.)}.

\subsection{Phantom Experiments}
We \textcolor{black}{acquired} image data of wave fields on gelatin phantoms that are manufactured 24 hours prior to data acquisition. The phantoms \textcolor{black}{were} removed from the \textcolor{black}{refrigerator} at least 2 hours prior to the experiments to limit variations of gelatin elasticities through temperature changes. We \textcolor{black}{manufactured} five homogeneous phantoms for each gelatin to water concentration of \SI{3}{\%}, \SI{5}{\%}, \SI{7.5}{\%}, \SI{10}{\%}, \SI{12.5}{\%} and \SI{15}{\%}. We strictly \textcolor{black}{followed} an in-house recipe: mix water and gelatin powder, soak for \SI{2}{\hour}, heat to \SI{45}{\degree} for \SI{10}{\minute}, add per \SI{100}{\milli\liter} solution \SI{0.7}{\gram} \textcolor{black}{\ch{TiO2}} and \SI{0.4}{\gram} graphite powder for speckle, and let the mixture cool down before pouring in a 3D printed mold. We \textcolor{black}{fabricated} cylindrical phantoms with a radius of \SI{10}{\milli\meter} and a height of \SI{40}{\milli\meter} and \textcolor{black}{performed} compression experiments with a high-resolution force sensor (Nano17, ATI, U.S.) attached to a robot (PI H-820, Physik Instrumente, Germany) and \textcolor{black}{estimated} the Young's modulus $E$, similar to~\cite{neidhardt2022ultrasound}. We strictly \textcolor{black}{followed} a protocol during indentations. Additionally, we \textcolor{black}{manufactured} gelatin phantoms that contain a stiff inclusion of \SI{12.5}{\%} embedded in softer \SI{7.5}{\%} gelatin, referred to as inclusion phantoms. The inclusion \textcolor{black}{had} a circular shape with a diameter of \SI{20}{\milli\meter}. After we \textcolor{black}{positioned} the piezoelectric element on the inclusion phantom we \textcolor{black}{acquired} an RGB image from the top and \textcolor{black}{segmented} the position of the excitation and the inclusion. This \textcolor{black}{allowed} us to generate ground truth elasticity maps, based on estimates from indentation experiments, and report a DICE score~\cite{zijdenbos1994} to assess the segmentation of a stiff inclusion.

\vspace{-1mm}
\subsection{Training and Evaluation}
To evaluate the performance of our model, we \textcolor{black}{performed} a nested five-fold cross-validation approach. We \textcolor{black}{trained} our network to predict the Young's Modulus which we \textcolor{black}{estimated} from indentation experiments~\cite{neidhardt2022ultrasound}. All trainings \textcolor{black}{were} performed with data acquired on homogenous gelatin phantoms. To reduce computational costs for our deep learning approach, we \textcolor{black}{performed} a bilinear interpolation along the lateral axis. We report a resolution of \qtyproduct{64 x 128 x 540}{pixels} for 2D+t and pseudo 3D+t data sets. We \textcolor{black}{acquired} training data on five phantoms of each gelatin concentration. On each phantom, 70 data sets at distinct positions \textcolor{black}{were} recorded with $d_s$ set to \SI{5}{\milli\meter}. Our training data set consists of over 350 million scan lines. For each fold, we \textcolor{black}{left} out one phantom of each gelatin concentration for testing (20$\%$). The remaining 4 phantoms \textcolor{black}{were} randomly split into 1 phantom for validation (20$\%$) and 3 phantoms (60$\%$) for training. We \textcolor{black}{trained} in total four networks for each fold while randomly choosing a phantom from each gelatin concentration for validation. For ST Maps we \textcolor{black}{performed} a bilinear interpolation leading to a size of \qtyproduct{64 x 1024}{pixels} along the lateral and time axis, respectively. For 3D data sets, we randomly \textcolor{black}{cropped} sequences along the temporal axis during training resulting in an input size of \qtyproduct{64 x 128 x 64}{pixels} for 2D+t and pseudo 3D+t data sets. Note that we \textcolor{black}{preserved} the full temporal resolution. We \textcolor{black}{trained} our network for 200 epochs with a batch size of 14 and a learning rate of 10e-5. We \textcolor{black}{used} Adam for optimization with the mean squared error as a loss function between our prediction and target labels. For a recorded data set we \textcolor{black}{estimated} the elasticity with our trained deep learning model by cropping 3D sequences along the temporal axis from 2D+t and pseudo 3D+t data sets. We shift a temporal window by increments of 8 time steps and \textcolor{black}{predicted} for each temporal window an elasticity. We report the mean of all estimates for a data set.

\subsection{Soft-Tissue Experiments}
We demonstrate our method on ex-vivo porcine soft tissue to show a clinical application for minimally invasive surgery on the heart. The porcine hearts were extracted less than 48 hours before data acquisition from a local butcher. The imaging was performed superficially on the left ventricle. We positioned the FOV so that it contains adipose as well as muscle tissue.

\begin{table*}[ht]
\caption{Results as absolute estimates for all experiments on homogeneous gelatin phantoms. All values are given in \SI[]{}{\kilo \pascal}.}


\centering
    \begin{tabular}{l l D{,}{\pm}{4.4} D{,}{\pm}{4.4} D{,}{\pm}{4.4} D{,}{\pm}{4.4} D{,}{\pm}{4.4} D{,}{\pm}{4.4} }
        \toprule
        \textit{Method}
        & \textit{Data}
        & \multicolumn{1}{c}{$G_{3\%}$}
        & \multicolumn{1}{c}{$G_{5\%}$}
        & \multicolumn{1}{c}{$G_{7.5\%}$}
        & \multicolumn{1}{c}{$G_{10\%}$}
        & \multicolumn{1}{c}{$G_{12.5\%}$}
        & \multicolumn{1}{c}{$G_{15\%}$}\\
        \midrule
        FFT 
        & ST Map
        & 28.12,27.10
        & 44.78,25.78
        & 56.63,24.43
        & 67.13,22.50
        & 84.47,19.71
        & 102.18,13.63\\
        FFT + AC
        & ST Map
        & 11.73,3.53
        & 24.84,4.64
        & 37.39,7.73
        & 45.86,9.63
        & 59.45,15.75
        & 74.81,24.62\\
        DL 
        & ST Map
        & 8.59,5.86 
        & 18.37,9.92 
        & 34.39,11.31 
        & 50.42,10.93 
        & 61.94,11.33 
        & 79.38,12.06 \\ 
        DL 
        & 2D+t
        & 14.76,9.54 
        & 18.71,6.83 
        & 42.59,12.63 
        & 57.10,9.90 
        & 66.16,11.11 
        & 89.13,9.53 \\  
        DL 
        & Pseudo 3D+t
        & 12.95,6.91 
        & 18.52,3.74 
        & 40.42,7.46 
        & 58.44,5.42 
        & 71.73,7.50 
        & 91.56,6.27\\
        \midrule
        \multicolumn{2}{r}{Indentation Test} 
        & \multicolumn{1}{c}{\SI{8.28}{\kilo\pascal}}
        & \multicolumn{1}{c}{\SI{17.42}{\kilo\pascal}}
        & \multicolumn{1}{c}{\SI{37.55}{\kilo\pascal}}
        & \multicolumn{1}{c}{\SI{56.04}{\kilo\pascal}}
        & \multicolumn{1}{c}{\SI{72.64}{\kilo\pascal}}
        & \multicolumn{1}{c}{\SI{97.22}{\kilo\pascal}}\\
        \bottomrule
    \end{tabular}    
    \label{tab:MAE_Table_1}
\end{table*}

\begin{table*}[ht]
\caption{Results as mean absolute error in kPa for all experiments on homogeneous gelatin phantoms.}


\centering
    \begin{tabular}{l l D{,}{\pm}{4.4} D{,}{\pm}{4.4} D{,}{\pm}{4.4} D{,}{\pm}{4.4} D{,}{\pm}{4.4} D{,}{\pm}{4.4} D{,}{\pm}{4.4}  }
        \toprule
\textit{Method} 
& \textit{Data} 
        & \multicolumn{1}{c}{$G_{3\%}$}
        & \multicolumn{1}{c}{$G_{5\%}$}
        & \multicolumn{1}{c}{$G_{7.5\%}$}
        & \multicolumn{1}{c}{$G_{10\%}$}
        & \multicolumn{1}{c}{$G_{12.5\%}$}
        & \multicolumn{1}{c}{$G_{15\%}$}
        & \multicolumn{1}{c}{Avg.}\\
     \midrule
FFT  & ST Map
& 19.84,27.10
& 27.39,25.75
& 20.78,23.00
& 17.67,17.79
& 17.43,14.96
& 12.13,79.1
& 19.75,21.82
\\

FFT + AC    & ST Map
& \textbf{3.49},\textbf{3.50}
& 7.59,4.37
& 6.12,4.71
& 11.48,8.03
& 15.99,12.89
& 25.82,21.00
& 11.72,13.23

\\  
DL  & ST Map
& 4.23,4.67 
& 7.31,6.88 
& 9.10,7.42 
& 9.70,7.54 
& 13.21,8.27 
& 18.49,11.04 
& 8.11,6.95 
\\
DL & 2D+t

& 8.09,8.27 
& 5.12,5.01 
& 10.41,8.75 
& 7.94,6.00 
& 10.13,7.92 
& 10.04,7.44 
& 6.31,5.76 
\\
DL & Pseudo 3D+t 
& 5.85,5.97 
& \textbf{2.61},\textbf{2.90 }
& \textbf{5.85},\textbf{5.45 }
& \textbf{4.54},\textbf{3.80 }
& \textbf{6.27},\textbf{4.21} 
& \textbf{6.74},\textbf{5.08} 
& \textbf{4.48},\textbf{3.63 }\\
        \bottomrule
    \end{tabular}    
    \label{tab:MAE_Table_2}
\end{table*}

\begin{figure*}[t]
    \includegraphics[width=\linewidth, trim={0 0 0cm 0},clip]{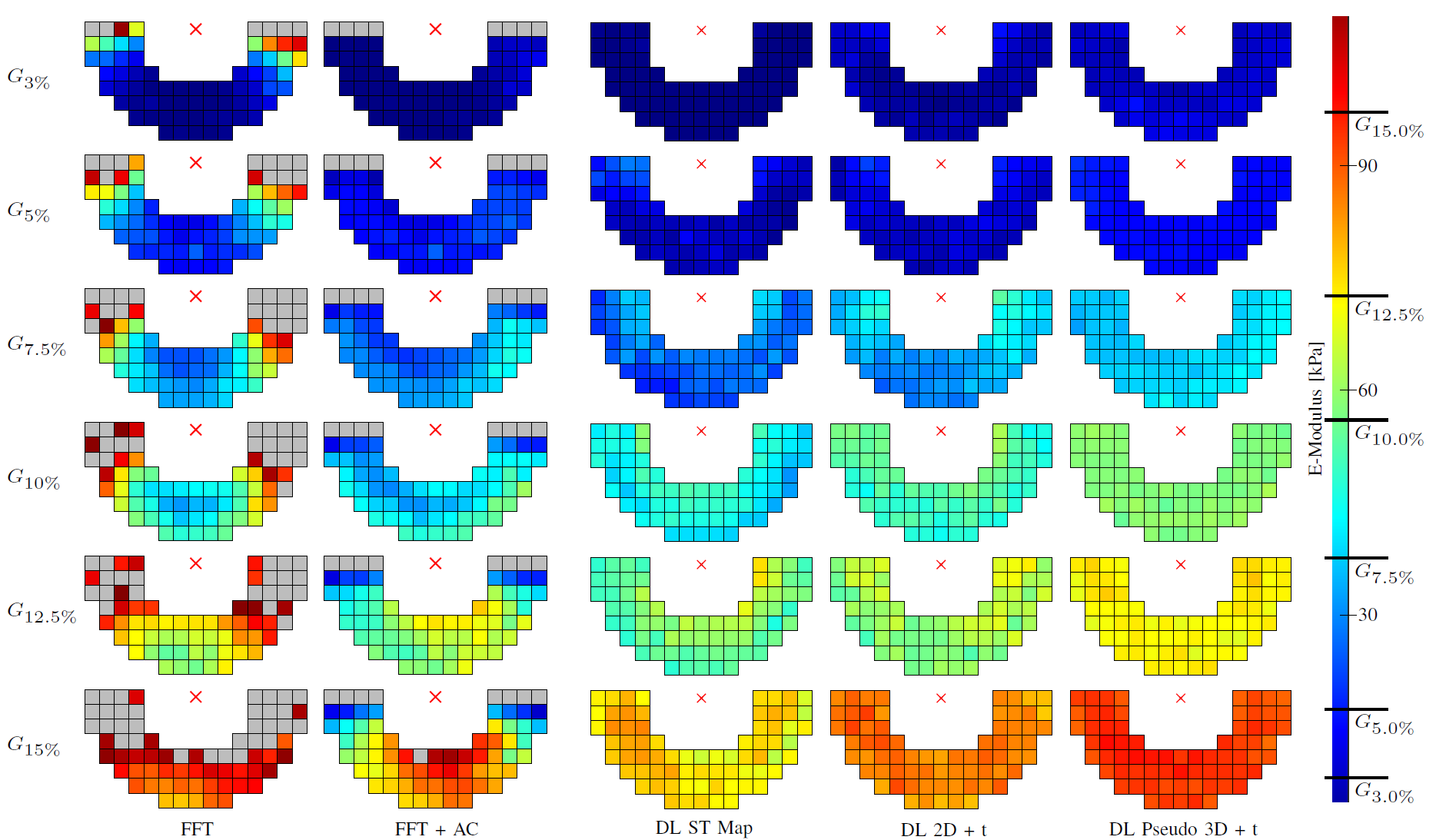};
    \caption{\textbf{\textcolor{black}{Elasticity} Maps of Gelatin Phantoms:} Each map represents the mean estimates for all seven phantoms of the individual gelatin concentrations. Estimates are given for the conventional method (columns 1-2) and for our deep learning (DL) approach (columns 3-5). The red cross indicates the excitation location and failed estimates are indicated in gray. The spatial dimension of each square is  \qtyproduct{5 x 5}{\milli\meter}.}
    \label{fig:HeatMaps}
\end{figure*}

\begin{figure*}[t]
    \includegraphics[width=\linewidth, trim={0 0 0cm 0},clip]{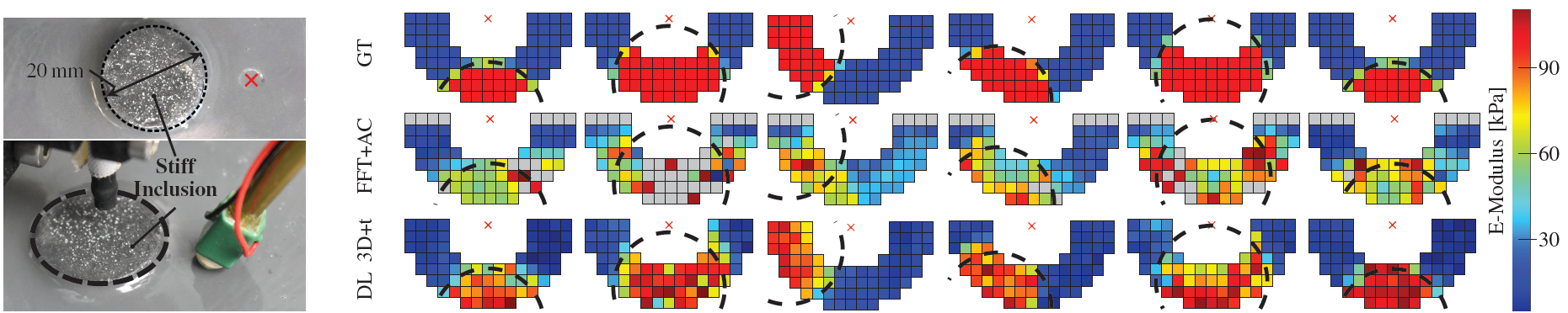};
    \caption{\textbf{Elasticity Maps of Inclusions} \textit{Left:} Phantom with stiff inclusion (r=\SI[]{10}{\milli\meter}). The top image is used for ground truth (GT) annotation. \textit{Right:} Elasticity maps of six individual phantoms. The excitation position is denoted as the red 'x'. Top row GT, center row predictions with conventional image processing (FFT + AC), and bottom row deep learning estimates from pseudo 3D+t image data. The dashed black line indicates the location of the inclusion. Failed estimates are indicated in gray. The spatial dimension of each square is \qtyproduct{2 x 2}{\milli\meter}.}
    \label{fig:InclusionMaps}
\end{figure*}

 \begin{figure}[t]
    \includegraphics[width=\linewidth, trim={0 0 0cm 0},clip]{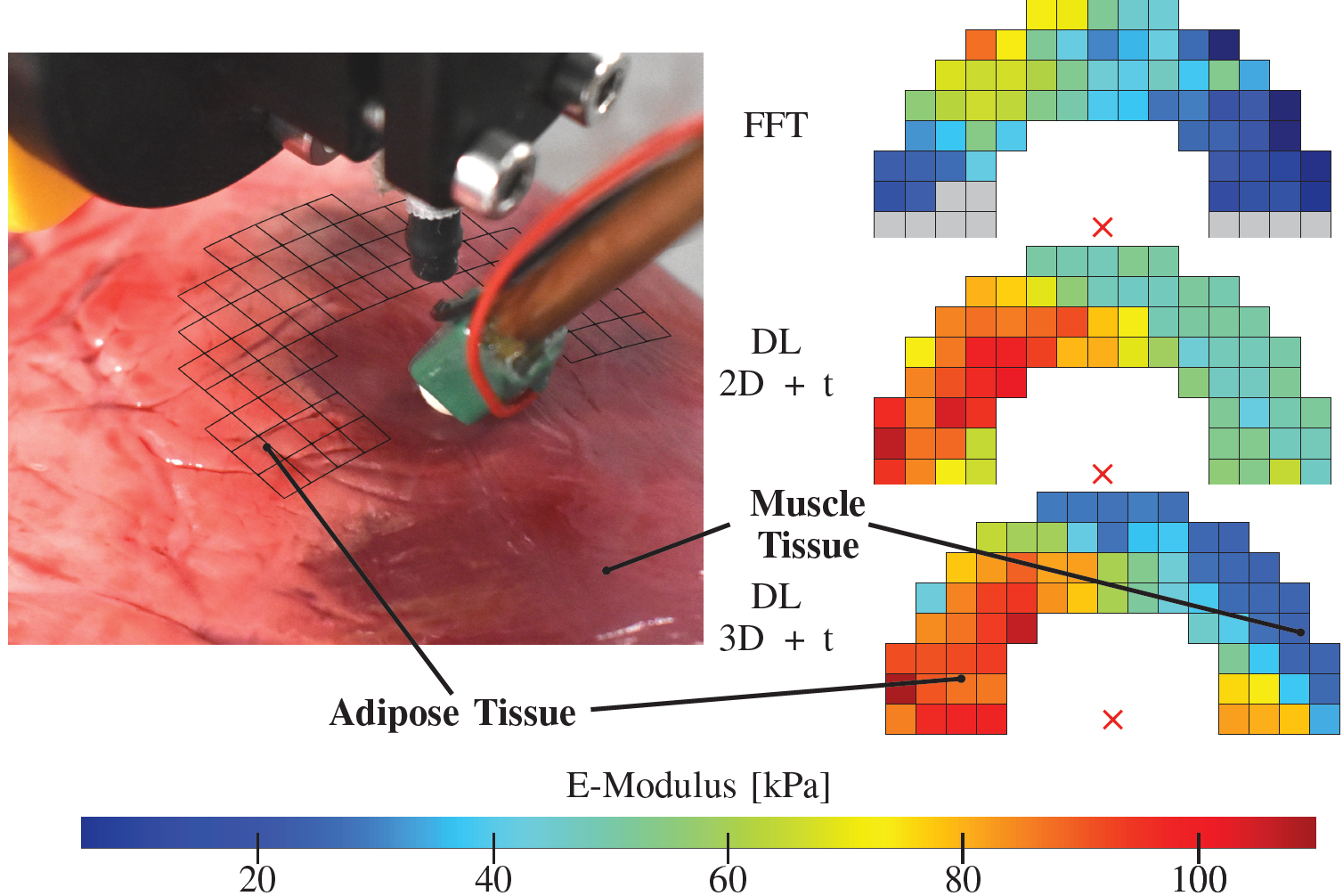};
    \caption{\textbf{Elasticity Maps of Pig Heart Tissue:} Left is depicted the experimental setup including the piezoelectric excitation element, the FSE driven by a robot (not depicted) and the outline of the scan field indicated in black. Right, are depicted elasticity maps obtained from the conventional approach (top), deep learning estimates from 2D+t (middle) and pseudo 3D+t scan mode (bottom). Failed estimates are indicated in gray. The spatial dimension of each square is \qtyproduct{2 x 2}{\milli\meter}.}
    \label{fig:InclusionMapsPig}
\end{figure}

\vspace{-3mm}

\section{Results} 
Table~\ref{tab:MAE_Table_1} shows the results for elasticity estimations performed on gelatin phantoms with a uniform elasticity. We estimate the absolute elasticity values in \SI{}{\kilo\pascal} and compare estimates to indentation tests performed on additionally manufactured cylindrical gelatin phantoms containing the same gelatin to water ratio. For our conventional elasticity estimation approach, we report a mean absolute error (MAE) of \SI{19.75(21.82)}{\kilo\pascal} for various positions on the phantom. When taking the angle of the wave into account (FFT + AC) the MAE is reduced to \SI{11.33(1278)}{\kilo\pascal}. The elasticity maps in Figure~\ref{fig:HeatMaps} indicate that the MAE increases for off-axis estimates. For softer gelatin phantoms and therefore slower propagating waves this error can be decreased if the angle of the wave is taken into account. However, for stiffer and therefore faster propagating waves the MAE is distinctively increased, even when taking the wave propagation direction into account. Moreover, estimates off-axis are predominantly removed and indicated as outliers in gray. In comparison, our deep learning approach outperforms the conventional elasticity estimations. We give the MAE for different input data types in Table~\ref{tab:MAE_Table_2}. For processing ST Map representations with deep learning, we report an MAE of \SI{8.11(0695)}{\kilo\pascal}, which is lower than the conventional elasticity estimation approaches. Still, image information along the depth axis is compromised during ST Map generation. In contrast, when processing 2D+t and pseudo 3D+t data representations we report an MAE of \SI{6.31(0576)}{\kilo\pascal} and \SI{4.48(0363)}{\kilo\pascal}, respectively. It stands out that apart from the $3\%$ gelatin concentration the MAE is always best when the FSE is operated in pseudo 3D+t scan mode. The elasticity maps in Figure~\ref{fig:HeatMaps} depict a uniform distribution of elasticity estimates. It stands out that our deep learning approach provides fair elasticity estimates at all positions on the gelatin phantom while the conventional method partially fails. Overall, the most uniform distribution of elasticity estimates is obtained when operating the endoscope in pseudo 3D+t scan mode.

Further, we evaluate the performance of estimating elasticity with the FSE on gelatin phantoms with an embedded stiff inclusion. Figure~\ref{fig:InclusionMaps} depicts the experimental setup on the left with the phantom, FSE, and piezoelectric element. Waves are excited at a fixed position on the phantom surface. Elasticity maps of six phantoms can be seen in Figure~\ref{fig:InclusionMaps}, right. The top row indicates the ground truth (GT) elasticity of the phantom with the respective gelatin concentrations. Elasticity estimates obtained with the conventional approach including the propagating direction (FFT + AC) are depicted in the center row and can be directly compared to elasticity estimates obtained with deep learning and the FSE operated in conical shape scan mode (deep learning pseudo 3D+t). It can be visually seen, that the distribution of estimates is more uniform while considering our deep learning approach in comparison to the conventional approach. Also, the deep learning approach estimates an elasticity value for each position while the conventional approach fails at various positions. Even for on-axis estimates the conventional method does not provide consistent estimates inside the stiff inclusion, as seen in phantom 2. For all phantoms we report a DICE score of \SI[]{0.64(10)} (precision: \SI[]{0.94(2.51)}) for elasticity estimates with FFT + AC and \SI[]{0.91(03)} (precision: \SI[]{0.82(11)}) with deep learning pseudo 3D+t. Elasticity maps obtained from soft tissue are depicted in Figure~\ref{fig:InclusionMapsPig}. While no ground truth is obtainable the results show qualitatively a distribution of elasticity estimates which corresponds to visual features on the soft tissue corresponding to adipose and muscle tissue. The trend in distribution is similar for conventional and deep learning based elasticity estimates.

\vspace{-3mm}

\section{Discussion} 

In this work, we present a deep learning approach utilizing a FSE for local elasticity estimation. The miniature design of our endoscope with an outer diameter of \SI[]{2.2}{\milli\meter} is ideal for integration into instruments used for minimally invasive surgery. To the best of our knowledge, this is the first application of a FSE for shear wave elasticity imaging. We acquire line or conical shape scan patterns with a megahertz OCT system during continuous wave excitation. The sparse scan patterns allow a higher temporal scan frequency which is beneficial for imaging rapidly changing wave fields. Miniature probes have been proposed that allow the acquisition of OCT images but are limited to one spatial dimension~\cite{Latus.2023, Parmar.2021, Karpiouk.2018, Qu.2017}. In contrast, with the presented FSE, we can acquire 2D+t or pseudo 3D+t images of the tissue sample with a temporal sampling rate of \SI{10.1}{\kilo\hertz} while estimating the elasticity of the tissue. By acquiring complete line or conical shape scan patterns, we do not rely on any spatial calibrations of the excitation location and the FOV, as depicted in Figure~\ref{fig:InclusionMaps} by varying the position of the stiff inclusion. Note, that the inclusion and soft-tissue phantom were not included in the training data set. Moreover, our data processing with deep learning does not rely on triggers and allows us to acquire image data at various positions and time points of a diffuse and complex wave field. Hence, the position of the FOV relative to the excitation location is not critical when acquiring three-dimensional scan sequences. This makes our approach highly flexible and simplifies our probe design by not needing to include the wave excitation inside the probe. Additionally, our small FOV of \SI{1.5}{\milli\meter} allows highly local elasticity estimates which is crucial for elasticity estimates on inhomogeneous and anisotropic tissues.

We demonstrate the performance of the FSE on tissue-mimicking phantoms. This allows us to analyze the wave propagation for different tissue stiffness in a reproducible and uniform fashion. We analyze the performance of our elasticity estimates on gelatin concentrations ranging from $3\%$ to $15\%$ which corresponds roughly to muscle tissue (\SIrange{12}{13}{\kilo \pascal})\cite{kot2012elastic} and tendon tissue (\SIrange{70}{72}{\kilo \pascal})\cite{kot2012elastic}, respectively. Our results show that deep learning estimates are more accurate than elasticity estimates performed with a conventional image processing approach. For estimates performed on $3\%$-gelatin phantoms with relatively slow wave propagation, the conventional method is slightly advantageous. Still, when analyzing the spatial relationship of excitation and imaging (Figure~\ref{fig:HeatMaps}) it can be seen that the conventional method fails when image data is acquired off-axis to the propagating waves. Also, for higher gelatin concentrations which correspond to faster propagating wave fields the accuracy is reduced. Taking into account the direction of wave propagation (FFT + AC), waves propagating perpendicular to the line scanning axis are not resolvable. Also, it needs to be considered that accurate tracking of probe and excitation location is crucial for this conventional approach, which is in general limited in a minimally invasive clinical environment.

For efficient data processing, we utilize a deep learning network to directly estimate tissue elasticity from image data. Commonly, conventional elasticity estimation is intensive in parameter tuning and strongly relies on an accurate material model. The most common models are based on the assumption of an isotropic, homogeneous, incompressible elastic solid. In contrast, our deep learning network is trained directly on uni-axial compression tests. This allows us to infer quantitative elasticity estimates directly in real time from image data. The moduli measured during the indentation experiment and shear wave elastography differ and are complex to model with conventional methods~\cite{chen2021comparative}. However, neural networks can be considered general function approximators which include the conversion to moduli obtained during indentation. For elasticity estimates from space times maps, which are the basis of conventional elasticity estimation approaches, we report a 28$\%$ improvement in elasticity estimation with our deep learning approach. Furthermore, when we increase the spatial information to 2D and 3D data structures, the accuracy is further enhanced by 44$\%$ and 60$\%$, respectively. Hence, the 3-dimensional information of waves propagating through the complex scan pattern is beneficial and even outperforms the 2D image data approach, although it provides a 2-fold temporal sampling rate in comparison. Processing of 3D image data is limited with conventional methods since an accurate and tedious spatial calibration of the scan field is necessary~\cite{SchulzHildebrandt.2018}. This calibration is commonly performed in air and changes depending on the underlying tissue. In contrast, our deep learning approach incorporates all spatial calibrations by systematically changing the positions of the probe relative to the wave excitation location for training data acquisition on phantoms. At each position, the wave propagates differently through the FOV, increasing data augmentation. In addition, we use a robot to acquire a large training data set. Also, imperfections of the scan field, e.g. by the in-house hand build probe, are considered during data inference and result in robust elasticity estimates. Our results show that the network can convert shear wave data to quantitative estimates based on our indentation tests. Additionally, deep learning offers the advantage of performing classification tasks directly, e.g., differentiating tissue types. Note, that in this case, end-to-end training would avoid indentation tests and instead use disease related labels.

The velocity of a surface or shear wave is dependent on the propagating depth inside the tissue and related to the excitation frequency~\cite{li2011elastic}. Surface waves with lower frequency penetrate deeper into the sample, while the higher frequency components are limited to shallower depths. Moreover, the velocity of the surface wave at different frequencies can be used for depth-resolved elasticity measurement in OCE~\cite{li2011elastic, li2012determining}. Accordingly, clinical research has shown that elasticity estimates are dependent on the excitation frequency due to the viscoelastic effects of soft-tissue~\cite{parker2018biomechanics, Hossain.2023}. This reinforces our approach of stimulating waves with multiple frequencies within the tissue. Additionally, this leads to increased data augmentation which is beneficial for deep learning approaches.

Real-time estimates of the elasticity with our deep learning based data processing approach are feasible. Image data is minimally pre-processed by surface detection and downsampling ($\sim$\SI{40}{\milli\second}). With an inference time of the network of \SI{22}{\milli\second} we report an approximate elasticity estimation rate of \SI{14}{Hz} including a subsidiary data processing time of \SI{10}{\milli\second}. This rate allows continuous scanning of surface structures. Motion artifacts can compromise the signal, however, the effect of typical motion would be small considering the high image frequency of up to \SI{10.1}{\kilo\hertz}. Also, maintaining the OCT FOV at the tissue surface will require image-based tracking, which has been demonstrated with a megahertz OCT system~\cite{schluter2020high}. In conjunction with our robot controller which operates at \SI{250}{Hz} even fast movements can be compensated. 

To enable clinical implementation, further in-vivo validation will be required and sterilization of the probe needs to be considered. Further, elasticity estimates on anisotropic mediums, e.g., muscle fibers, need to be studied, as the wave propagation characteristics depend on the underlying anisotropic tissue\cite{leong2020stiffness}. Multiple excitation sources could improve elasticity estimation in those cases. To limit the amount of surgical tools inside the patient the excitation of the tissue can be performed directly with a surgical instrument. This has been demonstrated by integrating piezoelectric elements inside a Da~Vinci robotic tool~\cite{neidhardt2024modified}. The size of our scan field already represents a typical laparoscopic image size of \qtyproduct{53 x 40}{\milli\meter} at a working distance of \SI{120}{\milli\meter}~\cite{qin2014characterization}. Further, interpretation of the elasticity and grayscale values can be performed by adapting the learning task. This leverages our approach to be highly flexible considering the surgical task.


\vspace{-1mm}

\section{Conclusion}
We present a novel 3D scanning probe and a deep learning based signal processing approach allowing for precise and real-time localized elasticity estimates independent of the wave propagation direction. Considering the contactless sensing and ability to move the miniaturized probe, we demonstrate feasibility to obtain elasticity maps with a high resolution at tissue surfaces. We illustrate that the approach can be used to quantify tissue elastic properties and the shape of stiffer regions, making it promising for aiding surgical navigation.

\vspace{-0.5mm}

\bibliographystyle{IEEEtran}
\bibliography{refs}

\end{document}